\documentclass[aps,prl,twocolumn,showpacs,floatfix]{revtex4}

\usepackage{graphicx}
\usepackage{amssymb}

\begin{document}

\title{Unusual upper critical field of the ferromagnetic superconductor UCoGe}

\author{N. T. Huy}
\author{D. E. de Nijs}
\author{Y. K. Huang}
\author{A. de Visser}
\email{devisser@science.uva.nl}

\affiliation{Van der Waals - Zeeman Institute, University of
Amsterdam, Valckenierstraat~65, 1018 XE Amsterdam, The
Netherlands}

\date{\today}

\begin{abstract}
We report upper critical field $B_{c2}(T)$ measurements on a
single-crystalline sample of the ferromagnetic superconductor
UCoGe. $B_{c2}(0)$ obtained for fields applied along the
orthorhombic axes exceeds the Pauli limit for $B \parallel a,b$
and shows a strong anisotropy $B_{c2}^{a} \simeq B_{c2}^{b} \gg
B_{c2} ^{c}$. This provide evidence for an equal spin pairing
state and a superconducting gap function of axial symmetry with
point nodes along the $c$ axis, which is also the direction of the
uniaxial ferromagnetic moment $m_0 = 0.07~ \mu_{B}$. An unusual
curvature or kink is observed in the temperature variation of
$B_{c2}$, which possibly indicates UCoGe is a two-band
ferromagnetic superconductor.

\end{abstract}


\maketitle

Recently, a new member of the family of itinerant ferromagnetic
superconductors has been reported, UCoGe~\cite{Huy-PRL-2007}.
Itinerant ferromagnetic superconductors attract much attention,
because they form an exception to the general rule which says that
superconductivity (SC) and ferromagnetic (FM) order are mutually
exclusive: in the standard BCS theory magnetic interactions impede
phonon-mediated singlet superconductivity~\cite{Berk-PRL-1966}.
The coexistence of superconductivity and metallic ferromagnetism
therefore calls for an exotic explanation, which is offered by
spin fluctuation models: on the border of ferromagnetic order
critical magnetic fluctuations mediate superconductivity by
pairing the electrons in triplet
states~\cite{Fay-PRB-1980,Lonzarich-CUP-1997}. The ferromagnetic
superconductors known to date are UGe$_2$ (under
pressure)~\cite{Saxena-Nature-2000}, possibly UIr (under
pressure)~\cite{Akazawa-JPCM-2004}, URhGe~\cite{Aoki-Nature-2001}
and the newcomer UCoGe~\cite{Huy-PRL-2007}. The latter two
materials are of special interest, because superconductivity
occurs at ambient pressure, which provides a unique opportunity to
apply a wide range of experimental techniques for the
investigation of magnetically mediated superconductivity.
Moreover, since URhGe and UCoGe both crystallize in the
orthorhombic TiNiSi structure (space group
$P_{nma}$)~\cite{Lloret-PhDthesis-1988}, but have different
magnetic interaction strengths, systematics in the variation of
the SC properties may be studied.

The magnetic and SC parameters reported for UCoGe so far, have
been extracted from measurements on polycrystalline
samples~\cite{Huy-PRL-2007}. Magnetization measurements show that
UCoGe is a weak FM with a Curie temperature $T_{C}$= 3 K and a
small ordered moment $m_{0}$= 0.03 $\mu_B$. The itinerant nature
of the FM state is corroborated by the small value of the magnetic
entropy (0.3 \% of Rln2) associated with the magnetic transition.
The SC properties depend sensitively on the quality of the
samples. SC is observed with a resistive transition temperature
$T_{s}$ = 0.8 K for the best sample. Thermal expansion and
specific-heat measurements provide solid evidence for bulk
magnetism and superconductivity.

In this Letter we report the first results of magnetic and
transport measurements on single-crystalline UCoGe. Magnetization
measurements show that UCoGe is a uniaxial FM, with an ordered
moment $m_{0}$= 0.07 $\mu_B$ pointing along the orthorhombic $c$
axis. The suppression of the SC transition in a magnetic field $B$
was studied by resistivity measurements. The upper critical field
$B_{c2}(T)$ measured for $B$ applied along the orthorhombic axes
is not Pauli limited for $B
\parallel a,b$ and shows a large anisotropy. The observed anisotropy provides evidence
for a SC gap function of axial symmetry with point nodes along the
direction of the ordered moment. Such a SC gap function is in
agreement with recent order parameter calculations for
orthorhombic ferromagnetic spin-triplet
superconductors~\cite{Fomin-JETPLett-2001,Mineev-PRB-2002,Mineev-PRB-2004}.
A pronounced positive curvature or kink is observed in
$B_{c2}^{a,b}(T)$ which possibly indicates two-band
superconductivity~\cite{Fay-PRB-1980,Mineev-PRB-2004,Belitz-PRB-2004}.

A polycrystalline batch with nominal composition U$_{1.01}$CoGe
was prepared by arc melting the constituents (natural U 3N, Co 3N
and Ge 5N) in a water-cooled copper crucible under a high-purity
argon atmosphere. Next, a single-crystalline rod was pulled from
the melt using a modified Czochralski technique in a tri-arc
furnace under a high-purity argon atmosphere. Electron micro-probe
analysis confirmed the single-phase nature of the grown crystal.
Single-crystallinity was checked by X-ray Laue backscattering.
Samples for various measurements were cut by spark erosion.
Transport measurements on the as-grown samples show a residual
resistance ratio, $RRR=R(300$K$)/R(1$K$) \approx 5$, and rather
poor FM and SC properties. However, after an annealing procedure,
like applied for URhGe~\cite{Hardy-PRL-2005},
the $RRR$ increases to 30 and proper FM and SC transitions appear.

Magnetization measurements were carried out down to temperatures
of 2 K and magnetic fields up to 5 T in a SQUID magnetometer on
bar-shaped samples cut along the different crystallographic
directions. Corrections due to demagnetizing fields are neglected,
because the demagnetization factor is small ($N \approx 0.1$).
Four-point low-frequency ac-resistivity data were obtained using a
phase-sensitive bridge for $T$= 0.02-10~K on a bar-shaped sample
with the current along the $a$ axis and $RRR=30$. In Fig.~1 we
show the resistivity $\rho (T)$ measured for a current along the
$a$ axis. The magnetic transition is observed as a sharp kink at
$T_{C} = 2.8~$K. Below 2 K $\rho \propto T^2$ due to scattering at
magnons. Superconductivity appears below $0.6~K$ and the
transition width $\Delta T_{s} \simeq 0.1$ K. Thermal expansion
data taken on the same sample demonstrate superconductivity is a
bulk property~\cite{Gasparini-tobepublished}.

\begin{figure}
\includegraphics[width=6.5cm]{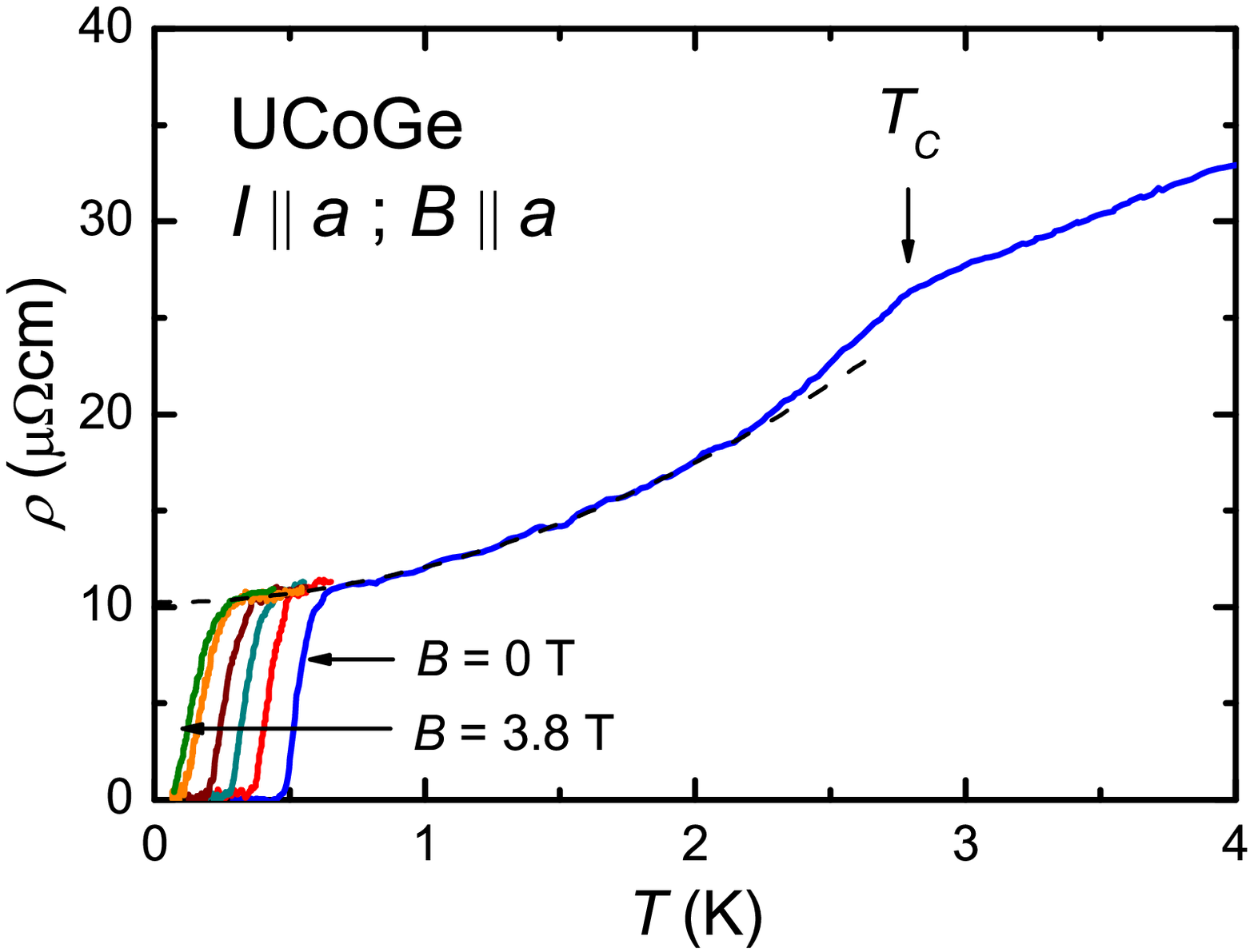}
\caption{Temperature variation of the resistivity of UCoGe for a
current $I \parallel a$. Ferromagnetic ordering at $T_{C} = 2.8$ K
shows up as a sharp kink. The dashed line represents $\rho (T) =
\rho_{0} + A T^2$ with $\rho_{0} = 10.2~ \mu \Omega$cm and $A=
1.8~ \mu \Omega$cmK$^{-2}$. Superconductivity appears below 0.6 K.
In a magnetic field ($B \parallel a$) superconductivity is
suppressed ($B$-values from right to left are 0, 0.8, 1.6, 2.4,
3.4 and 3.8 T).}
\end{figure}

The suppression of superconductivity was investigated by
resistivity measurements in fixed magnetic fields applied along
the orthorhombic $a$ (longitudinal), and $b$ and $c$ axis
(transverse). In a field $\Delta T_{s}$ gradually increases to
0.15 K at the highest fields (5~T). The temperature at which SC is
suppressed is taken by the mid-points of the transitions. In an
applied field the FM state rapidly forms a mono domain ($B < 0.01$
T) and we did not observe any hysteric behavior in $B_{c2}$. The
main results are shown in Fig.~2. At least three remarkable
features appear in the data: (i) the large value of $B_{c2}(0)
\approx 5$ T for $B
\parallel a,b$, (ii) the large anisotropy, $B_{c2}^{a} \simeq B_{c2}^{b} \gg B_{c2} ^{c}$,
of a factor $\sim 10$, and (iii) for all $B$-directions
$B_{c2}(T)$ has a pronounced upturn when lowering the temperature.
Clearly, this behavior is at odds with standard BCS spin-singlet
pairing.

\begin{figure}
\includegraphics[width=6.5cm]{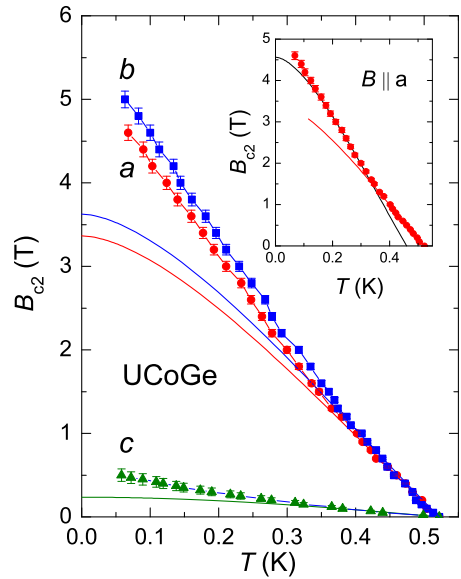}
\caption{Temperature dependence of the upper critical field of
UCoGe for $B$ along the $a$, $b$ and $c$ axis. The solid lines
show the calculated dependence for a superconducting gap function
with axial (along $c$) and polar symmetries (along $a$ and $b$)
(see text). Inset: $B_{c2}(T)$ for $B
\parallel a$, plotted together with the polar state functions for
two superconducting phases with zero-field transitions at 0.46 and
0.52 K. }
\end{figure}

Let us first address the large $B_{c2}^{a,b}(0)$. The Pauli
paramagnetic limit for a weak coupling spin-singlet superconductor
is $B_{c2}^{Pauli}(0) = 1.83 T_s$~\cite{Clogston-PRL-1962}. By
including the effect of spin-orbit coupling a comparable
$B_{c2}(0)$ value~\cite{Werthamer-PR-1966} results. Combined Pauli
and orbital limiting could therefore in principle account for the
small value of $B_{c2}^{c}(0)$, but not for the large
$B_{c2}^{a,b}(0)$ values. Also a strong-coupling scenario is
unrealistic as this would involve a huge coupling constant
$\lambda \approx 20$~\cite{Rainer-PRB-1973}. The absence of Pauli
limiting therefore points to a triplet SC state with equal-spin
pairing (ESP). However, a prerequisite for triplet pairing is a
sufficiently clean sample, with a mean free path $\ell$ larger
than the coherence length $\xi$~\cite{Millis-PRB-1988}. An
estimate for $\ell$ and $\xi$ can be extracted within a simple
model~\cite{Orlando-PRB-1979} from the (large) initial slope
$dB_{c2}^{a,b}/dT$ = - 7.9-8.4 T/K. With the specific heat
coefficient  $\gamma$ = 0.057 J/molK$^2$~\cite{Huy-PRL-2007} and
$\rho_{0} = 10.2~ \mu \Omega$cm we calculate $\ell \approx
900$~\AA~and~$\xi \approx 120$~\AA~and conclude our single crystal
satisfies the clean-limit condition.

The order parameter in an orthorhombic itinerant ferromagnetic
superconductor with spin-triplet pairing and strong spin-orbit
coupling has been worked out by Fomin~\cite{Fomin-JETPLett-2001}
and Mineev~\cite{Mineev-PRB-2002}. Under the assumption that the
exchange splitting of the Fermi surface is sufficiently large,
such that the pairing between electrons from spin-up and down
bands is negligible, equal-spin pairing will give rise to two-band
superconductivity with gap functions $\Delta_{\uparrow
\uparrow}(R,k) = - \eta _{1}(R) f_{-}(k)$ and $\Delta_{\downarrow
\downarrow}(R,k) =  \eta _{2}(R) f_{+}(k)$, where $\eta _{1,2}$
are the order parameter amplitudes. From the symmetry-group
analysis for an orthorhombic uniaxial FM it follows that only two
SC gap-structures $f_{\pm}(k)$ are possible. Assuming that the
ordered moment $m_{0}$ is directed along the $z$ axis, then the SC
gap has zeros (nodes) parallel to the magnetic axis
($k_{x}=k_{y}=0$, $A$ phase) or a line of zeros on the equator of
the Fermi surface ($k_{z} =0$, $B$ phase). In other words, the $A$
phase has a gap function of axial symmetry with nodes along $m_0$
and the $B$ phase has a gap of polar symmetry with a line of nodes
perpendicular to $m_0$. Before relating the anisotropy in $B_{c2}$
to the structure of the SC gap we first examine the magnetization
$M(B)$.

$M(B)$ for a field along the $a$, $b$ and $c$ axis is presented in
Fig.~3. The data reveal UCoGe is an uniaxial FM with the ordered
moment $m_{0} \parallel c$. The inset shows $M_{c}(T)$ measured in
a small field $B=0.01$~T. By smoothly extrapolating the data for
$T \rightarrow 0$ we obtain $m_{0} = 0.07 ~\mu_{B}$. This value is
in agreement with the powder averaged value $m_{0}^{powder} =
0.03~\mu_{B} \simeq m_{0}^{c-axis} /2$~\cite{Huy-PRL-2007}. For $B
\parallel a,b$ the induced magnetization is small. The large
initial increase in $M_{c}(B)$ is related to the relatively high
temperature of ~0.7$T_{C}$ at which the data are taken.

\begin{figure}
\includegraphics[width=6.5cm]{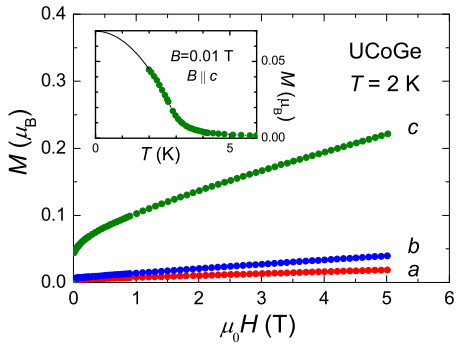}
\caption{Magnetization of UCoGe for fields along the $a$, $b$ and
$c$ axis at $T=$ 2 K. Ferromagnetic order is uniaxial with $m_{0}$
pointing along the $c$ axis. The inset shows $M(T)$ for $B$ = 0.01
T along $c$. In the limit $T \rightarrow 0~ m_{0} = 0.07~
\mu_{B}$. }
\end{figure}

Having established that UCoGe is a uniaxial ferromagnet with
$m_{0} \parallel c$, we now turn to the anisotropy in $B_{c2}$.
The effect of an anisotropic $p$-wave interaction on $B_{c2}$ has
been investigated by Scharnberg and
Klemm~\cite{Scharnberg-PRL-1985}. For a $p$-wave interaction that
favors one direction over the other two, a polar state has the
highest $T_s$, while conversely, if the $p$-wave interaction is
weakest in one direction an axial state is favored. Since the SC
gap is fixed to the crystal lattice by spin-orbit coupling,
$B_{c2}$ in general will show a strong anisotropy. We first
consider the case of a polar state with the maximum gap direction
along the uniaxial direction $m_{0}
\parallel c$ ({\it i.e.} the strength of the
pairing interaction along $c$ is stronger than in the $ab$ plane,
$V_{c} > V_{ab}$). The upper critical fields have been calculated
for $B$ in the plane of the nodes ($B \perp m_{0}$) (CBS =
Completely Broken Symmetry state) and $B \parallel m_{0}$ along
the maximum of the gap (polar state)~\cite{Scharnberg-PRL-1985}.
For $V_{ab}=0$ their ratio for $T \rightarrow 0$ is given by
$B_{c2}^{\perp}(0)/B_{c2}^{\parallel}(0)=
0.466(m_{ab}/m_{c})^{1/2}$, where $m_{ab}/m_{c}$ reflects the
anisotropy in the effective mass. For our orthorhombic material
$m_{ab}/m_{c}$ is of the order of 1 and the model predicts
$B_{c2}^{\perp}(0)/B_{c2}^{\parallel}(0)$ is $\sim 1/2$. This is
at variance with the results in Fig.~2 where the anisotropy ratio
$\sim 10$ and we conclude that a polar gap cannot explain the
anisotropy in $B_{c2}$. For the axial state (nodes along $m_0
\parallel c$ and the maximum gap in the $ab$ plane) the anisotropy
in $B_{c2}$ is reversed~\cite{Scharnberg-PRL-1985}
$B_{c2}^{\perp}(0) > B_{c2}^{\parallel}(0)$. This is the situation
in UCoGe. $B_{c2}^{\parallel}(T)$ has been calculated in
Ref.~\cite{Scharnberg-PRB-1980} (ABM state), but calculations for
$B_{c2}^{\perp}(T)$ are not at hand. However, since the gap is
maximum one may assume that $B_{c2}^{\perp}(T)$ can be represented
by the polar function. In Fig.~2 we compare the anisotropy in
$B_{c2}$ calculated in this way with the experimental results. We
conclude that the measured anisotropy supports an axial state, but
the model calculations do not track the experimental results at
lower $T$. This is due to the pronounced positive curvature in
$B_{c2}(T)$, which we discuss next.

For $B \parallel a$ and $b$ the slope -$dB_{c2}/dT$ has initial
values of 7.9 and 8.4 T/K but increases to 11.4 and 12.1 T/K at
lower $T$. For $B \parallel a$ the increase is rather abrupt
(kink-like) and takes place in a narrow $T$ range centered around
0.33 K (see inset in Fig.~2), while for $B
\parallel b$ the increase is more gradual.
An overall smooth increase of $B_{c2}$ is observed for $B
\parallel c$. We stress that if we define $T_s$ in another way,
like by $T_{s}^{onset}$ or by the 10\% or 90 \% points (measured
by the drop in resistance) the absolute values for $B_{c2}$ change
slightly, but the upward curvature remains. An upward curvature in
$B_{c2}(T)$ was also reported for the polycrystalline
samples~\cite{Huy-PRL-2007}.

Several appealing explanations for a kink ($B \parallel a$) or
upward curvature ($B \parallel b$) in $B_{c2}(T)$ have been given
in the literature, of which a cross-over between two phases in a
two-band ferromagnetic superconductor~\cite{Mineev-PRB-2004} is
perhaps the most appealing. Mineev and
Champel~\cite{Mineev-PRB-2004} have evaluated the linearized
Ginzburg-Landau equations including gradients terms for a cubic
two-band FM superconductor with gaps $\Delta_{\uparrow \uparrow}$
and $\Delta_{\downarrow \downarrow}$. Depending on the strength of
the pairing interactions measured by $g_{1} = V_{\uparrow
\uparrow} \langle |f_{-}(k)|^2 N_{0\uparrow}(k) \rangle$ and
$g_{2} = V_{\downarrow \downarrow} \langle|f_{+}(k)|^2 N_{0
\downarrow}(k) \rangle$ (here $N_{0 \downarrow,\uparrow}$ is the
angular dependent density of states) and a number of anisotropy
coefficients, an upturn in $B_{c2}(T)$ is predicted. In the more
anisotropic situation of an orthorhombic system, calculations are
more tedious, but also then a cross-over between ESP states in
field is possible. Notice that in the two-band model it is not
{\it a priori} known what the ground state is ($\mid \uparrow
\uparrow \rangle $ or $\mid \downarrow \downarrow \rangle$), as
this depends on $g_{1}$ and $g_{2}$ and the anisotropy
coefficients. In order to illustrate the possibility of the
two-band superconductivity scenario we have plotted in the inset
in Fig.~2 the polar state functions~\cite{Scharnberg-PRB-1980} for
two SC phases with zero-field transition temperatures of 0.46 and
0.52 K. A description with two polar gaps tracks the experimental
data down to $T \simeq 0.16$ K, but fails there below. Note
however that in the two-band model of Ref.~\cite{Mineev-PRB-2004}
in zero field only one superconducting transition occurs and that
the upturn in $B_{c2}$ is due to a field-induced crossover. This
is in accordance with the zero-field thermal expansion data, which
show a single SC transition~\cite{Gasparini-tobepublished}.
Another interesting scenario for an upwards curvature in $B_{c2}$
is a rotation of the quantization axis of the ESP state from along
$m_0$ at low fields, towards the field direction in high fields.
The upward curvature in $B_{c2}^{c}$ is then possibly explained by
the field not being perfectly aligned along $c$. Scharnberg and
Klemm~\cite{Scharnberg-PRL-1985} have also investigated the
possibility of a kink in $B_{c2}$. For the axial symmetric case a
kink appears for $B \parallel c$ for a specific ratio of the
strength of the pairing interaction, namely $V_{c}/V_{ab}
 < 0.866$, which could give rise to a transition from the Scharnberg-Klemm state
to the polar state. However, an upturn is not predicted for $B$
directed in the $ab$ plane. The only other unconventional
superconductor which shows a clear kink in $B_{c2}$ is the
heavy-fermion material UPt$_3$~\cite{Rauchschwalbe-ZPhysB-1985}.
Here the kink is attributed to the intersection of two $B_{c2}(T)$
curves of the two SC phases~\cite{Hess-JPCM-1989} even in the
absence of a magnetic field. The split transition is attributed to
the coupling of a two-component order parameter, belonging to the
$E_{1g}$ or $E_{2u}$ representation in the hexagonal crystal, to a
symmetry breaking field.

The upper critical field of the FM superconductor URhGe
(isostructural to UCoGe) has been investigated by Hardy and
Huxley~\cite{Hardy-PRL-2005}. $B_{c2}$, measured on single
crystals with $RRR$'s of 21 and 34, {\it i.e.} comparable to the
value of the UCoGe single-crystal, also show a large anisotropy.
The data do not show any sign of an upward curvature and are well
described by the model functions for a polar state with a maximum
gap parallel to $a$, and for the CBS
state~\cite{Scharnberg-PRL-1985} with nodes in the direction of
$b$ and $c$. This polar gap structure takes into account the
easy-($bc$)-plane nature of the magnetization: $m_{0} = 0.42~ \mu
_{B}$ points along $c$, but for $B \parallel b$ rotates towards
the $b$ axis~\cite{Levy-Science-2005}. Surprisingly,
superconductivity re-appears when the component of the applied
field along the $b$ axis reaches $\sim 12$
T~\cite{Levy-Science-2005} and the ordered moment rotates towards
the $b$ axis. Compelling evidence is at hand that in the
high-field as well as in the low-field SC phase, superconductivity
is stimulated by critical magnetic fluctuations associated with
the field-induced spin-reorientation process. Near the quantum
critical point an acute enhancement of $B_{c2}$ is
observed~\cite{Levy-NaturePhysics-2007} and superconductivity
survives in fields as large as 28 T applied along $a$. It will be
highly interesting to investigate whether the apparent lack of
saturation of $B_{c2}$ for $T \rightarrow 0$ in UCoGe has a
similar origin, namely the proximity to a field-induced quantum
critical point.

In conclusion, we have reported the first measurements of the
magnetic and superconducting parameters of a single crystal of
UCoGe. We find that ferromagnetic order is uniaxial with the
ordered moment $m_0 = 0.07~ \mu_{B}$ pointing along the $c$ axis.
The upper critical field for $B \perp m_0$ exceeds the upper
critical field for $B \parallel m_0$ by a factor 10. The magnitude
and anisotropy of $B_{c2}$ support $p$-wave superconductivity and
point to an axial SC gap function with nodes along the moment
direction ($A$ phase~\cite{Mineev-PRB-2002}). For $B \perp m_0$ we
observe a pronounced upward curvature ($B
\parallel b$) or kink ($B \parallel a$). An appealing explanation
for this unusual phenomenon is that UCoGe is a two-band
ferromagnetic
superconductor~\cite{Fay-PRB-1980,Mineev-PRB-2004,Belitz-PRB-2004}.
The structure in $B_{c2}$ could then mark a cross-over between two
SC phases with equal-spin pairing states. On the other hand, UCoGe
is close to a ferromagnetic instability as the Curie temperature
is low and $m_0$ is small. This brings to the fore another exotic
explanation for the positive curvature of $B_{c2}(T)$, namely
field-tuning of critical magnetic fluctuations, which promote and
enhance magnetically mediated superconductivity.

This work was part of the research program of FOM (Dutch
Foundation for Fundamental Research of Matter) and COST Action P16
ECOM. We acknowledge helpful discussions with M. Baranov and A.
Huxley.

\end{document}